\documentclass[prd,reprint,showpacs,showkeys]{revtex4-1}
\usepackage{amsfonts}
\usepackage{amssymb}
\usepackage{amsmath}
\usepackage{graphicx}
\usepackage{float}
\usepackage{subcaption}
\usepackage[font={footnotesize,it}]{caption}

\begin{document}

\title{Note on a thin-shell wormhole in extremal Reissner-Nordstr\"{o}m
geometry}
\author{S. Habib Mazharimousavi}
\email{habib.mazhari@emu.edu.tr}
\author{M. Halilsoy}
\email{mustafa.halilsoy@emu.edu.tr}
\affiliation{Department of Physics, Eastern Mediterranean University, Gazima\u{g}usa,
North Cyprus, Mersin 10 - Turkey.}

\begin{abstract}
We show that the cold horizon of the extremal Reissner-Nordstr\"{o}m can be
considered as the throat of a thin-shell wormhole with zero total exotic
matter and positive angular pressure. Such a wormhole is physical and stable
against radial perturbations provided an appropriate perfect fluid exists at
the throat.
\end{abstract}

\maketitle

\section{Introduction}

The idea of thin-shell wormholes \cite{WH} was developed in order to confine
the exotic matter sources encountered in wormholes to a narrow hypersurface
so that the vast bulk spacetime still possess physical (non-exotic) sources.
The induced metric on the throat caries a surface energy-momentum of fluid
type which satisfies the junction conditions due to Israel's junction
formalism \cite{Israel}. Such a surface fluid is taken to be of the form $%
S_{\mu }^{\nu }=diag\left( -\sigma ,p,p\right) $ in which $\sigma $ stands
for the energy-density on the shell and $p$ is the angular pressure. The
choice of equation of state for the fluid is problematic since it doesn't
satisfy all requirements. It is preferable to have $\sigma >0$ and a
reasonable equation of state i.e., $p=p\left( \sigma \right) $ so that the
energy conditions, at least the null and weak ones are satisfied. In the
past, for instance, we had given some models of thin-shell wormholes with
non-spherical geometries which had the local energy density negative ($%
\sigma <0$) but overall integration of $\sigma $ yields a positive total
energy \cite{MH}. Restriction of the induced metric to a spherical topology $%
R\times S^{2}$, unfortunately dashes much of the good aspects in this line
of thought. Yet, within the spherical topology it is our belief that there
are possible ways to construct physical, i.e. non-exotic thin-shell
wormholes \cite{MH}. In brief, this is the main purpose of the present
article. By choosing as our bulk spacetime the extremal Reissner-Nordstr\"{o}%
m (ERN) geometry we show that the energy-density on the shell vanishes with
a surface pressure $p=\frac{2}{M},$ which is inversely proportional to the
mass (or charge) of the ERN black hole. Vanishing of the energy density ($%
\sigma =0$) at the static equilibrium of the thin-shell wormhole is an
advantage since to say the least it is better than being negative ($\sigma
<0 $). In addition to the energy aspect, stability condition of the
thin-shell against linear radial perturbations is of utmost importance. The
fact that the shell must be stable against such perturbations for physical
viability makes the problem of thin-shell wormholes further difficult to
tackle. In this regard we perturb the thin-shell so that the radius becomes
a function of the proper time $R\left( \tau \right) $. Employing a variable
equation of state \cite{VEOS1,VEOS2} for the fluid on the shell yields,
together with the conservation equation, appropriate forms for $\sigma $ and 
$p$ which reduce correctly to their static (unperturbed) limits. In \cite%
{VEOS2} where the variable equation of state has been introduced, Varela
resolved the anomaly of the instability of the thin-shell wormhole in
Schwarzschild bulk at $a_{0}=3M$ i.e., the throat's radius equal to the
radius of photon sphere \cite{SCTS}.

As we shall consider the ERN to be the bulk metric we would like to mention
that Reissner-Nordstr\"{o}m thin-shell wormhole has been considered before,
first in \cite{RNTS1} and later in \cite{RNTS2}. In \cite{RNTS1} the
equation of state was set to $\frac{dp}{d\sigma }=\beta ^{2}\left( \sigma
\right) $ and in their numerical analysis they considered $\beta
_{0}^{2}=\beta ^{2}\left( \sigma _{0}\right) .$ Also in \cite{RNTS2} where
the equation of state was set to be a Chaplygin gas, $a_{0}=r_{h}$ has been
excluded due to the non-physical zone, as they called it in which no
solution exists.

\section{The formalism}

In the standard method of constructing a thin-shell wormhole in spherically
symmetric bulk \cite{VEOS1, SSTS} we have%
\begin{equation}
ds_{bulk}^{2}=-f\left( r\right) dt^{2}+\frac{dr^{2}}{f\left( r\right) }%
+r^{2}\left( d\theta ^{2}+\sin ^{2}\theta d\phi ^{2}\right)
\end{equation}%
with induced metric at the throat 
\begin{equation}
ds_{throat}^{2}=-d\tau ^{2}+R^{2}\left( \tau \right) \left( d\theta
^{2}+\sin ^{2}\theta d\phi ^{2}\right)
\end{equation}%
in which $\tau $ stands for the proper time. We find the energy density and
angular pressure from $S_{\mu }^{\nu }=diag\left( -\sigma ,p,p\right) $ of
the perfect fluid presented at the throat given by ($8\pi G=1$)%
\begin{equation}
\sigma =-4\left( \frac{\sqrt{f\left( R\right) +\dot{R}^{2}}}{R\left( \tau
\right) }\right)
\end{equation}%
and%
\begin{equation}
p=2\left( \frac{2\ddot{R}\left( \tau \right) +f^{\prime }\left( R\right) }{2%
\sqrt{f\left( R\right) +\dot{R}^{2}}}+\frac{\sqrt{f\left( R\right) +\dot{R}%
^{2}}}{R\left( \tau \right) }\right) .
\end{equation}%
Note that we use the notation such that a dot stands for $\frac{d}{d\tau }$
and a prime implies $\frac{d}{dR}$. In static condition one finds%
\begin{equation}
\sigma _{0}=-\frac{4\sqrt{f\left( R_{0}\right) }}{R_{0}},
\end{equation}%
and%
\begin{equation}
p_{0}=2\left( \frac{f^{\prime }\left( R_{0}\right) }{2\sqrt{f\left(
R_{0}\right) }}+\frac{\sqrt{f\left( R_{0}\right) }}{R_{0}}\right) .
\end{equation}%
Next, if we assume that the bulk spacetime is a black hole with the event
horizon at $r=r_{h}$, naturally one must impose $R_{0}>r_{h}$. This is due
to the fact that $p_{0}$ is diverging at the horizon which is not acceptable
for a physical wormhole. Note that $R_{0}=r_{h}$ is also excluded for the
same reason.

\subsection{Thin-shell wormhole in extremal black hole bulks}

Here we consider the bulk spacetime to be an extremal black hole such that $%
f\left( r_{h}\right) =0$ and is same for its first derivative i.e., $%
f^{\prime }\left( r_{h}\right) =0.$ More precisely we set $f\left( r\right)
=U\left( r\right) ^{2}$ with $U\left( r_{h}\right) =0$ satisfying $U\left(
r\right) >0$ for $r>r_{h}.$ In this configuration if we set $R_{0}=r_{h}$ we
find $\sigma _{0}=0$ and 
\begin{equation}
p_{0}=2U^{\prime }\left( r_{h}\right) 
\end{equation}%
which is positive and finite. The well known extremal Reissner-Nordstr\"{o}m
(ERN) black hole is a candidate for this arrangement. The metric of ERN is
given by $f\left( r\right) =\left( 1-\frac{M}{r}\right) ^{2}$ with a double
root event horizon located at $r_{h}=M$ ($=Q=$the charge). Hence the angular
pressure of the thin-shell wormhole located at $R_{0}=M$ is given by $p_{0}=%
\frac{2}{M}$ with vanishing energy density $\sigma _{0}=0$. For the static
setting in the frame of comoving observer located on the throat with proper
time $\tau ,$ the thin-shell is made of a perfect fluid with the energy
density zero and a positive angular pressure $p_{0}=\frac{2}{M}$ which is
inversely proportional to the mass of the ERN black hole.

\subsection{Stability}

Our next concern is to investigate the stability of the thin-shell wormhole
with a throat located at the horizon of an extremal black hole,
specifically, ERN. We perturb the thin-shell wormhole radially and due to
that the energy density and pressure become dynamic given by%
\begin{equation}
\sigma =-4\left( \frac{\sqrt{U^{2}+\dot{R}^{2}}}{R\left( \tau \right) }%
\right)
\end{equation}%
and%
\begin{equation}
p=2\left( \frac{\ddot{R}\left( \tau \right) +UU^{\prime }}{\sqrt{U^{2}+\dot{R%
}^{2}}}+\frac{\sqrt{U^{2}+\dot{R}^{2}}}{R\left( \tau \right) }\right)
\end{equation}%
where $R\left( \tau \right) $ is the throat radius after the perturbation.
One can show that $\sigma $ and $p$ satisfy the energy conservation law
given by%
\begin{equation}
\frac{d\sigma }{d\tau }=-\frac{2\dot{R}}{R}\left( p+\sigma \right)
\end{equation}%
or equivalently%
\begin{equation}
\frac{d\sigma }{dR}=-\frac{2}{R}\left( p+\sigma \right) .
\end{equation}%
%
%
%
%
%
%
%
%
%
%
%
%
%
%
%
%
%
%
%
%
%
%%%%%%%%%%%%%%%%%%%%%%%%%%%%%%%%%%%%%%%%%%%%%%%%%%
\begin{figure}[h]
\includegraphics[width=70mm,scale=0.7]{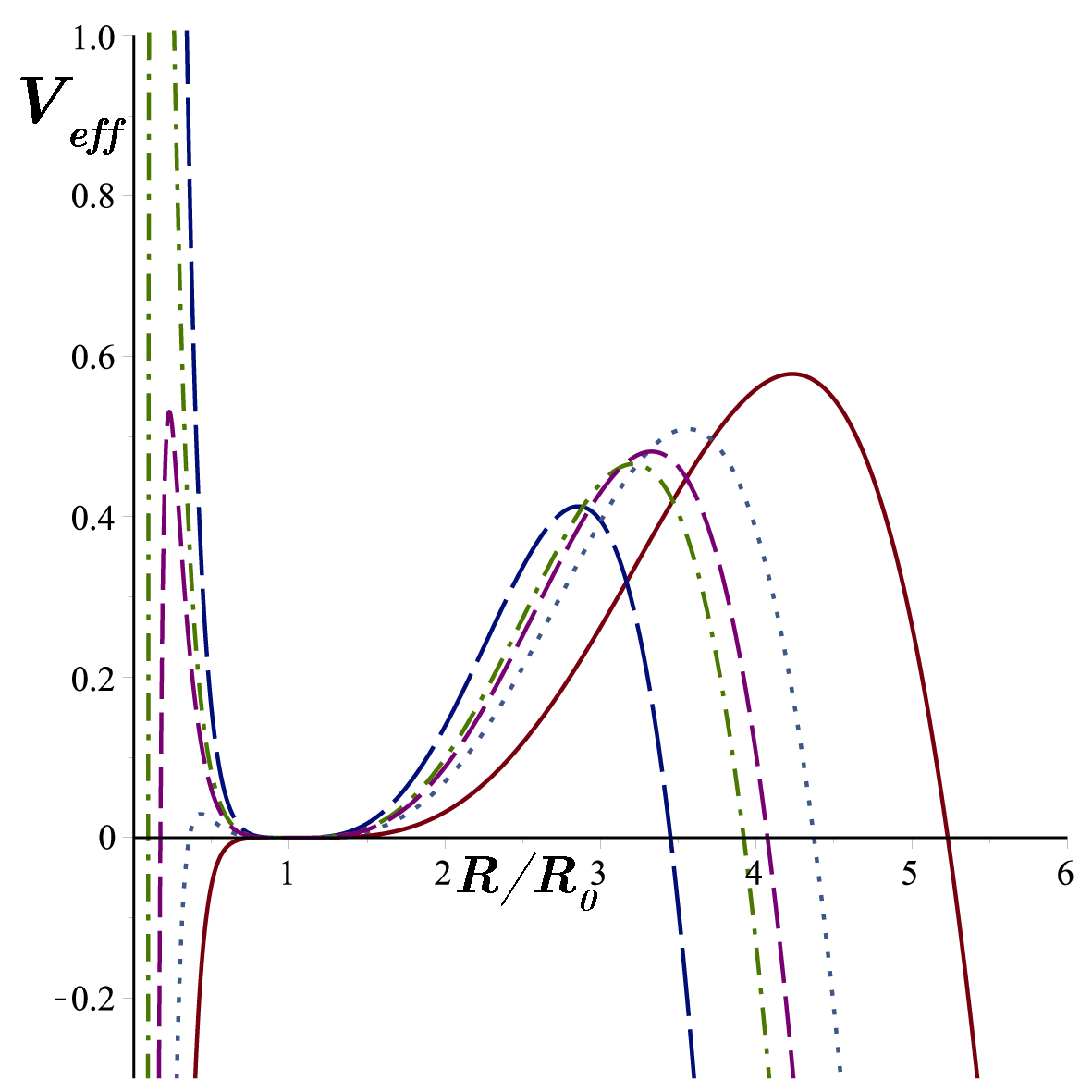}
\caption{Effective potential $V_{eff}$ vs $\frac{R}{R_{0}}$ for $\protect%
\omega =0,\frac{1}{10},\frac{1}{8},\frac{1}{6}$ and $\frac{1}{4}$ in
long-dash, dot-dash,dash, dot and solid respectively.}
\end{figure}

\begin{figure}[h]
\includegraphics[width=70mm,scale=0.7]{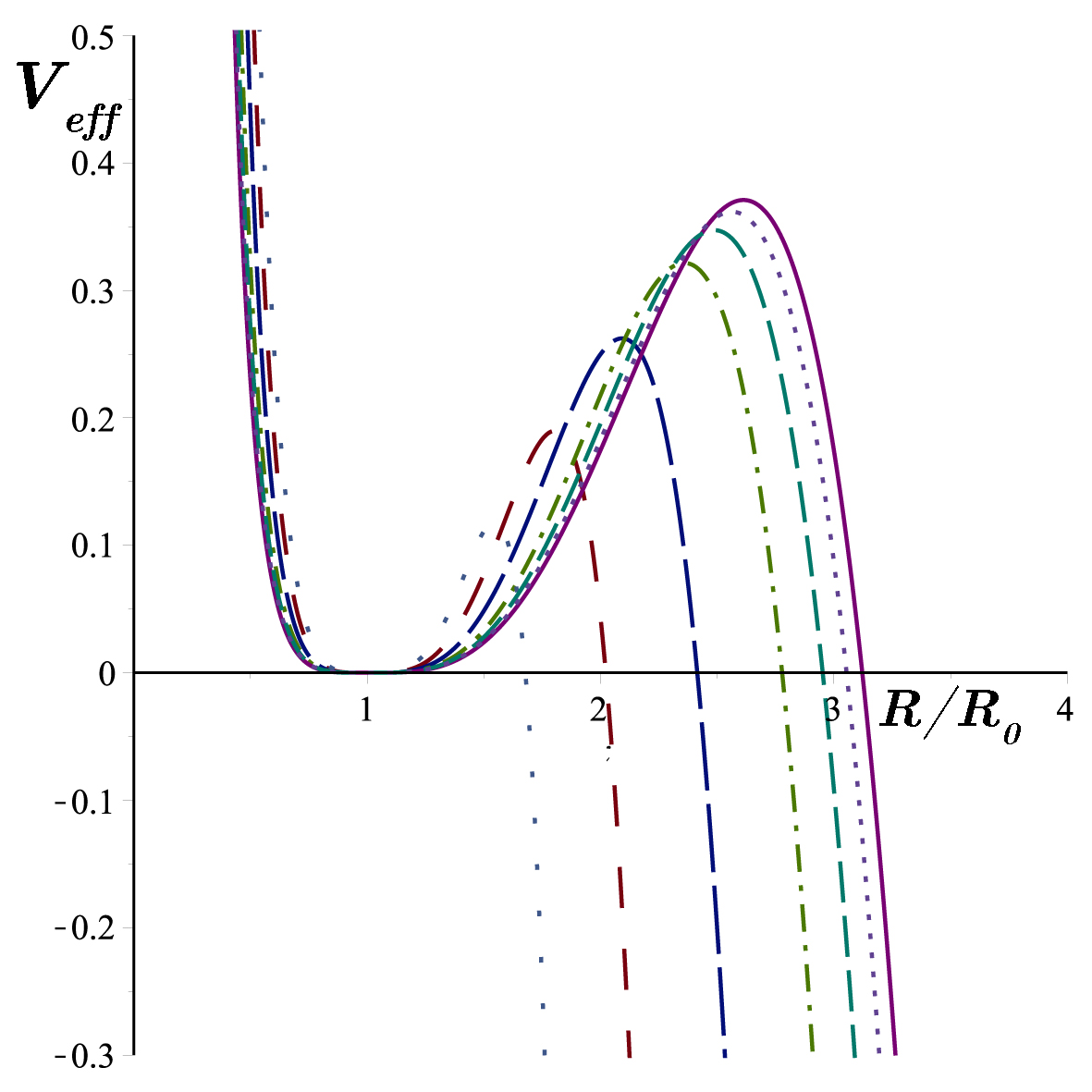}
\caption{Effective potential $V_{eff}$ vs $\frac{R}{R_{0}}$ for $\protect%
\omega =-\frac{1}{10},-\frac{1}{8},-\frac{1}{6},-\frac{1}{4},-\frac{1}{2},-1$
and $-2$ in solid, dot, dash, dot-dash,long-dash, space-dash and space-dot
respectively. In the case of $\protect\omega =-1$ one must use the limit of
the potential.}
\end{figure}
%%%%%%%%%%%%%%%%%%%%%%%%%%%%%%%%%%%%%%%%%%%%%%%%%%%%%
In addition to that one also finds from (3) that%
\begin{equation}
\dot{R}^{2}+\left( U^{2}-\left( \frac{\sigma R}{4}\right) ^{2}\right) =0
\end{equation}%
which is the equation of motion of the shell with an effective potential $%
V_{eff}=U^{2}-\left( \frac{\sigma R}{4}\right) ^{2}$. The last two equations
are coupled such that one can find a solution for $\sigma $ in (11) and
consequently from (12) we find the behaviour of $R$ with respect to the
proper time $\tau .$ To find $\sigma $ from (11) we must assume an equation
of state of the form $p=p\left( \sigma \right) $ which satisfies the
boundary conditions i.e., $p\left( \sigma _{0}=0\right) =p_{0}$ at $R=R_{0}.$
After \cite{VEOS1,VEOS2} we choose the following equation of state for the
perfect fluid at the location of the throat%
\begin{equation}
p=\omega \sigma +\frac{2\left( 2\omega -1\right) }{R_{0}}\left( \frac{R}{%
R_{0}}-1\right) +\frac{2}{R_{0}}
\end{equation}%
in which $\omega $ is a constant. Hence, the conservation equation admits%
\begin{equation}
\sigma =\frac{4\left( \frac{R}{R_{0}}-3\right) -4\omega \left( 2\omega
+1\right) \left( \frac{R}{R_{0}}-1\right) +8\left( \frac{R_{0}}{R}\right)
^{2\left( \omega +1\right) }}{R_{0}\left( \omega +1\right) \left( 3+2\omega
\right) }
\end{equation}%
in which the integration constant has been set such that $\sigma \left(
R_{0}\right) =0$ and naturally $\omega =-\frac{3}{2}$ must be excluded. The
case for $\omega =-1$ can be checked to exist from the L'Hospital's limiting
procedure. Considering this energy density in the effective potential of the
one-dimensional equation of motion (12) we find 
\begin{multline}
V_{eff}\left( R\right) =\left( 1-\frac{R_{0}}{R}\right) ^{2}- \\
\left( \frac{\omega -1}{\omega +1}\frac{R}{R_{0}}+\frac{2\left( \frac{R_{0}}{%
R}\right) ^{2\omega +1}}{\left( \omega +1\right) \left( 3+2\omega \right) }-%
\frac{2\omega -1}{3+2\omega }\left( \frac{R}{R_{0}}\right) ^{2}\right) ^{2}.
\end{multline}%
%
%
%
%
%
%
%
%
%
%
%
%
%
%
%
%
%
%
%
%
%
%%%%%%%%%%%%%%%%%%%%%%%%%%%%%%%%%%%%%%%%%%%%%%%%%%
\begin{figure}[h]
\includegraphics[width=70mm,scale=0.7]{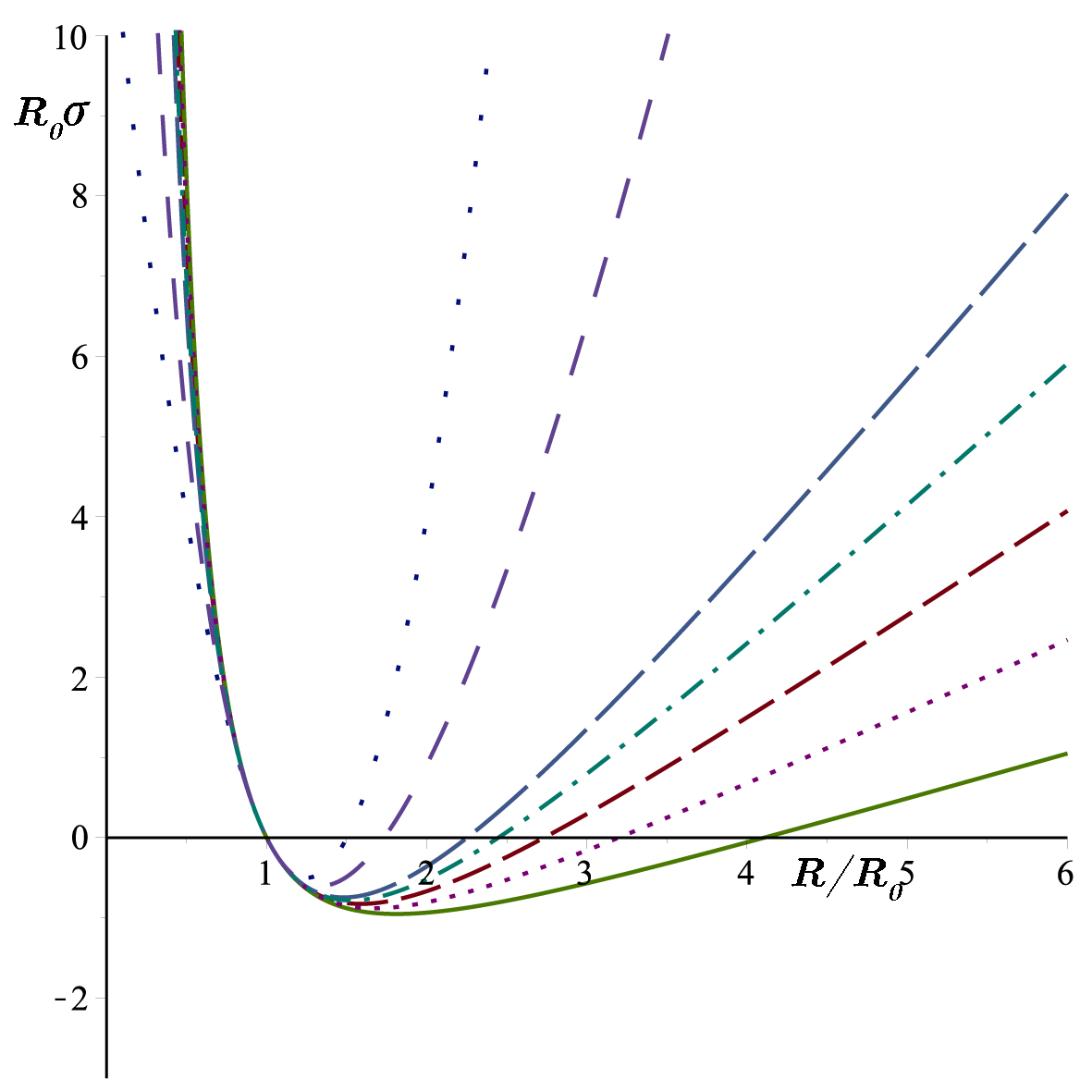}
\caption{$R_{0}\protect\sigma $ vs $\frac{R}{R_{0}}$ for $\protect\omega =%
\frac{1}{4},\frac{1}{8},0,-\frac{1}{8},-\frac{1}{4},-1$ and $-2$ in solid,
dot, dash, dot-dash,long-dash, space-dash and space-dot respectively.}
\end{figure}

\begin{figure}[h]
\includegraphics[width=70mm,scale=0.7]{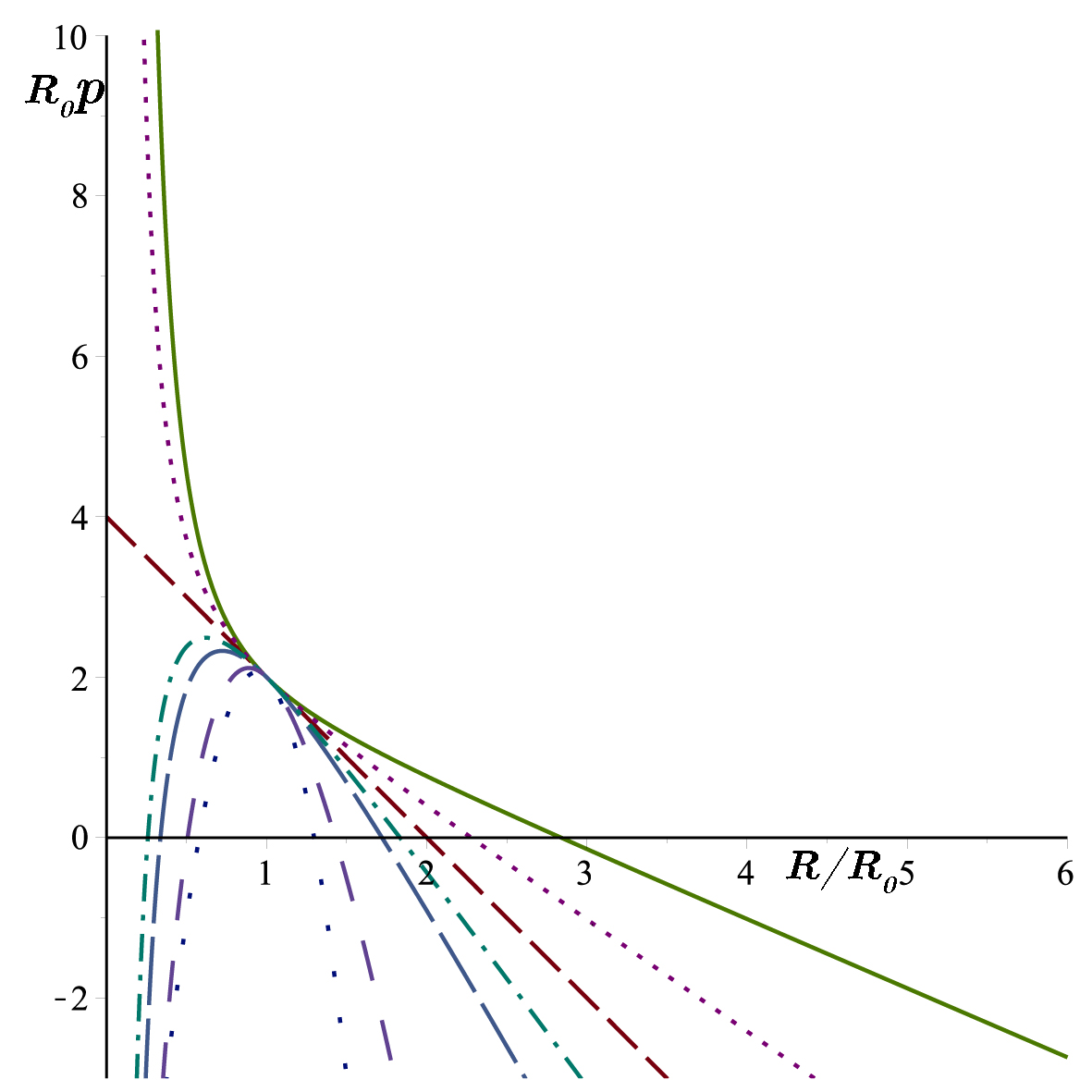}
\caption{$R_{0}p$ vs $\frac{R}{R_{0}}$ for $\protect\omega =\frac{1}{4},%
\frac{1}{8},0,-\frac{1}{8},-\frac{1}{4},-1$ and $-2$ in solid, dot, dash,
dot-dash,long-dash, space-dash and space-dot respectively.}
\end{figure}
\begin{figure}[h]
\includegraphics[width=70mm,scale=0.7]{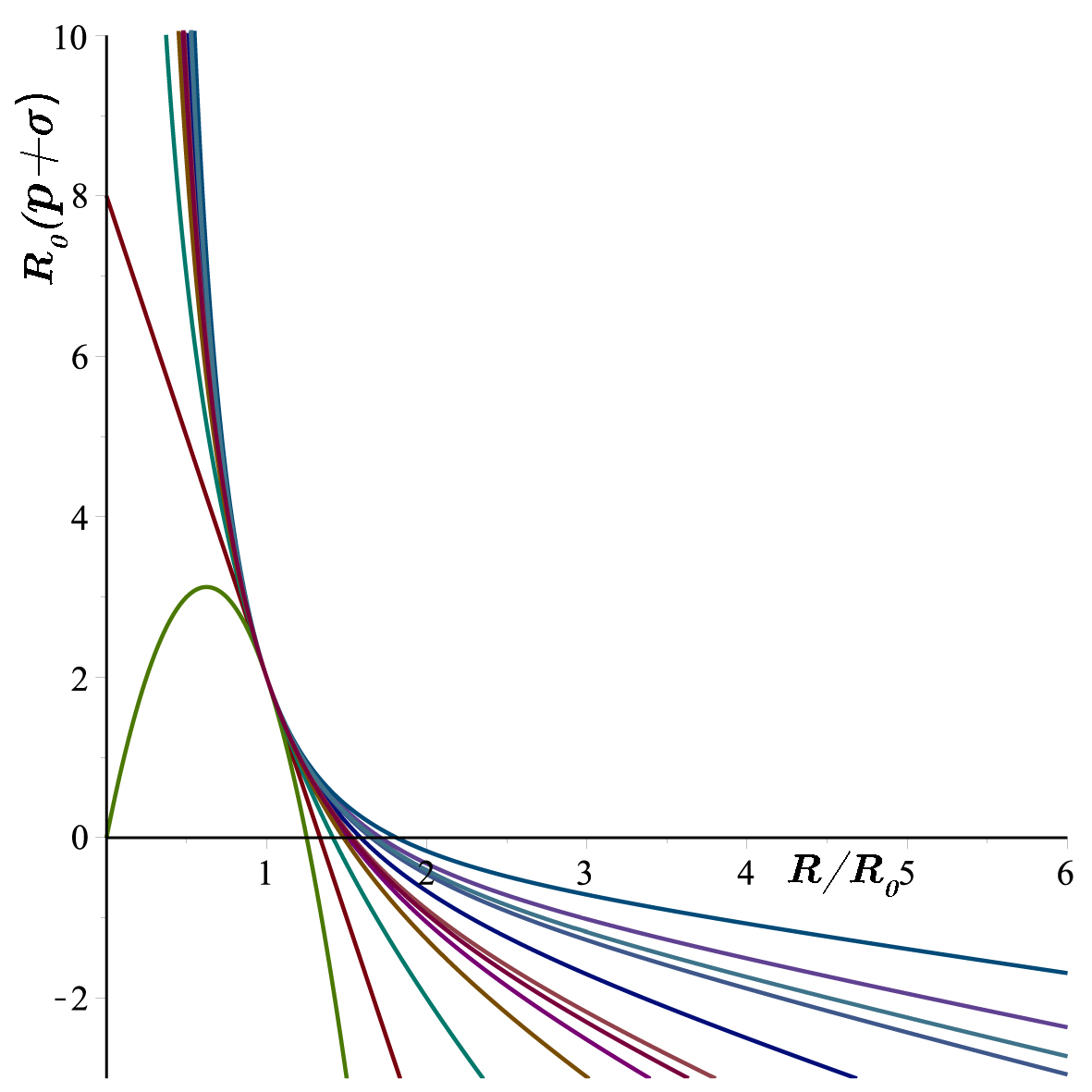}
\caption{$R_{0}\left( p+\protect\sigma \right) $ vs $\frac{R}{R_{0}}$ for $%
\protect\omega =\frac{1}{4},\frac{1}{6},\frac{1}{8},\frac{1}{10},0,-\frac{1}{%
10},-\frac{1}{8},-\frac{1}{4},-\frac{1}{2},-1$ and $-2$ from the right to
the left respectively.}
\end{figure}
%%%%%%%%%%%%%%%%%%%%%%%%%%%%%%%%%%%%%%%%%%%%%%%%%%%%%
In Figs. 1 and 2 we plot $V_{eff}\left( R\right) $ versus $\frac{R}{R_{0}}$
for various values of $\omega <\frac{1}{4}.$ Admitting a relative minimum at 
$\frac{R}{R_{0}}=1$ implies that the thin-shell wormhole is stable against
radial perturbations. As is clear from the Figs. 1 and 2, in the
neighborhood of $\omega =0$ the stability is stronger as the heights of the
barriers are getting higher. Furthermore, for strong perturbation even with $%
\omega <\frac{1}{4}$ the thin-shell may be unstable while for positive $%
\omega $ both possibility i.e., collapse and evaporation are likely to
happen whereas for $\omega <0$ the collapse can not take place. We also
observe that for $\omega \geq \frac{1}{4}$ the potential does not admit a
minimum at the location of the throat $R=R_{0}$ and therefore the thin-shell
wormhole becomes unstable. In Figs. 3 and 4 we plot $R_{0}\sigma $ and $%
R_{0}p$ in terms of $\frac{R}{R_{0}}$ for some values of $\omega $ which
have been used in Figs. 1 and 2. In Fig. 5 we plot $R_{0}\left( \sigma
+p\right) $ versus $\frac{R}{R_{0}}$ for various $\omega $'s$.$ Clearly in
the vicinity of $\frac{R}{R_{0}}=1,$ $\sigma +p$ is positive, indicating
that the energy conditions are satisfied, at least for $\frac{R}{R_{0}}\leq
1 $.

\section{Conclusion}

We constructed thin-shell wormholes in extremal Reissner-Nordstr\"{o}m
spacetime with its static throat $R=R_{0}$ located at the double root
horizon i.e., $r_{h}=R_{0}=M=Q$. We have shown that at static equilibrium
the perfect fluid presented at the throat possesses $\sigma _{0}=0$ and $%
p_{0}=\frac{2}{M}$ which clearly satisfy the energy conditions. Within the
equation of energy conservation (11) we found the exact forms of the energy
density and angular pressure when the thin-shell is perturbed radially. Also
we found a one-dimensional equation of motion for the throat after the
perturbation whose one-dimensional effective potential admits (for specific
values of $\omega $) a local minimum at the location of the static radius.
This is an indication for the stability of the thin-shell against a radial
perturbation. The equation of state which we adopted on the shell is of the
form $p=\omega \sigma +\frac{2\left( 2\omega -1\right) }{R_{0}}\left( \frac{R%
}{R_{0}}-1\right) +\frac{2}{R_{0}},$ in which $\omega $ is a constant
parameter. The fact that the role of the parameter $\omega $ is critical
both in stability and energy conditions can be seen from the plotted curves.
The perturbation equation for the radius of the shell satisfies an equation $%
\dot{R}^{2}+V_{eff}=0,$ where $\dot{R}=\frac{dR}{d\tau }$ and $V_{eff}$
stands for an effective potential i.e., Eq. (12). The plot of the potential $%
V_{eff}$ versus $\frac{R}{R_{0}}$ yields potential well in the vicinity of $%
\frac{R}{R_{0}}=1$ which renders stable configurations for finely tuned
parameter $\omega .$ We plot also $\sigma \left( R\right) $ which yields
positive values in certain regions of $R$. Further to $\sigma ,$ we also
investigate the combination $\sigma +p$, which possesses a positive domain
and therefore satisfies the energy conditions. Beyond certain range for the
parameter $\omega $ we observe that the shell becomes unstable. As a result
there are two possibilities: either the shell collapses to the center or it
expands indefinitely. As a final point let us remark that what has been done
about the cold, i.e., zero Hawking temperature, ERN black hole, are also
valid for other cold / ultracold extremal black holes.


\begin{thebibliography}{99}
\bibitem{WH} M. Visser, Phys. Rev. D \textbf{39}, 3182 (1989).

\bibitem{Israel} G. Darmois (1927) M\'{e}morial de Sciences Math\'{e}%
matiques, Fascicule XXV, \textquotedblleft Les equations de la gravitation
einsteinienne\textquotedblright , Chapitre V.

W. Israel "Thin shells in general relativity". Il. Nuovo Cim. \textbf{66}, 1
(1966).

\bibitem{MH} S. H. Mazharimousavi and M. Halilsoy, Eur. Phys. J. C \textbf{75%
}, 540 (2015);

S. H. Mazharimousavi and M. Halilsoy, Eur. Phys. J. C \textbf{75}, 271
(2015);

S. H. Mazharimousavi and M. Halilsoy, Eur. Phys. J. C \textbf{75}, 81 (2015);

S. H. Mazharimousavi and M. Halilsoy, Eur. Phys. J. C \textbf{74}, 3067
(2014).

\bibitem{VEOS1} N.\thinspace M. Garcia, F.\thinspace S.\thinspace N. Lobo
and M. Visser, Phys. Rev. D \textbf{86}, 044026 (2012).

\bibitem{VEOS2} V. Varela, Phys. Rev. D \textbf{92}, 044002 (2015).

\bibitem{SCTS} E. Poisson and M. Visser, Phys. Rev. D \textbf{52}, 7318
(1995);

F.\thinspace S.\thinspace N. Lobo and P. Crawford, Classical Quantum Gravity 
\textbf{21}, 391 (2004);

G.\thinspace A.\thinspace S. Dias and J.\thinspace P.\thinspace S. Lemos,
Phys. Rev. D \textbf{82}, 084023 (2010).

\bibitem{SSTS} E. F. Eiroa, Phys. Rev. D \textbf{78}, 024018 (2008).

\bibitem{RNTS1} E. F. Eiroa and G. E. Romero, Gen. Relativ. Gravit. \textbf{%
36}, 651 (2004).

\bibitem{RNTS2} M. Sharif and M. Azam, Eur. Phys. J. C \textbf{73}, 2554
(2013).
\end{thebibliography}
\end{document}